\documentclass[10pt,apm,amssymb,twocolumn,tightenlines,superscriptaddress,showpacs,floatfix]{revtex4}

\usepackage{graphicx}
\usepackage{times}
\usepackage{amssymb}
\usepackage{mathrsfs}
\usepackage{amsmath,amsfonts,latexsym}
\usepackage{array,tabularx,color}
\usepackage{dcolumn}  % Align table columns on decimal point
\usepackage[normalem]{ulem}
\usepackage{comment}
\usepackage{bm}
\usepackage{wasysym}
\usepackage{soul}

\begin{document}
\title{Spin Selective Filtering of Polariton Condensate Flow}
%Polariton Spinor Optical Field Effect Transistor

\author{T. Gao}
\affiliation{FORTH-IESL, P.O. Box 1385, 71110 Heraklion, Crete, Greece}
\affiliation{Department of Materials Science and Technology, Univ. of Crete, 71003 Heraklion, Crete, Greece}

\author{C. Ant\'on}
\affiliation{Departamento de F\'isica de Materiales, Universidad Aut\'onoma de Madrid, Madrid 28049, Spain}
\affiliation{Instituto de Ciencia de Materiales ``Nicol\'as Cabrera'', Universidad Aut\'onoma de Madrid, Madrid 28049, Spain}

\author{T. C. H. Liew}
\affiliation{School of Physical and Mathematical Sciences, Nanyang Technological University, 637371, Singapore}

\author{M. D. Mart\'in}
\affiliation{Departamento de F\'isica de Materiales, Universidad Aut\'onoma de Madrid, Madrid 28049, Spain}
\affiliation{Instituto de Ciencia de Materiales ``Nicol\'as Cabrera'', Universidad Aut\'onoma de Madrid, Madrid 28049, Spain}

\author{Z. Hatzopoulos}
\affiliation{FORTH-IESL, P.O. Box 1385, 71110 Heraklion, Crete, Greece}
\affiliation{Department of Physics, University of Crete, 71003 Heraklion, Crete, Greece}

\author{L. Vi\~na}
\affiliation{Departamento de F\'isica de Materiales, Universidad Aut\'onoma de Madrid, Madrid 28049, Spain}
\affiliation{Instituto de Ciencia de Materiales ``Nicol\'as Cabrera'', Universidad Aut\'onoma de Madrid, Madrid 28049, Spain}
\affiliation{Instituto de F\'isica de la Materia Condensada, Universidad Aut\'onoma de Madrid, Madrid 28049, Spain}

\author{P. S. Eldridge}
\affiliation{FORTH-IESL, P.O. Box 1385, 71110 Heraklion, Crete, Greece}
\affiliation{Department of Chemistry and Biochemistry, University of Delaware, 19716, United States}

\author{P. G. Savvidis}
\email{psav@materials.uoc.gr}
\affiliation{FORTH-IESL, P.O. Box 1385, 71110 Heraklion, Crete, Greece}
\affiliation{Department of Materials Science and Technology, Univ. of Crete, 71003 Heraklion, Crete, Greece}
\affiliation{Cavendish Laboratory, University of Cambridge, Cambridge CB3 0HE, United Kingdom}

\begin{abstract}
Spin-selective spatial filtering of propagating polariton condensates, using a controllable spin-dependent gating barrier, in a one-dimensional semiconductor microcavity ridge waveguide is reported. A nonresonant laser beam provides the source of propagating polaritons while a second circularly polarized weak beam imprints a spin dependent potential barrier, which gates the polariton flow and generates polariton spin currents. A complete spin-based control over the blocked and transmitted polaritons is obtained by varying the gate polarization.
\end{abstract}
 \maketitle

The generation of spin currents from the passage of electrons through ferromagnets is an important building-block for spintronic devices. For example, magnetic random access memory has developed from the spin valve concept, in which the current through a pair of ferromagnets is strongly attenuated when the magnets have opposite orientations~\cite{Chappert2007}. This principle has a striking analogy in optics, where the attenuation of light passing through crossed polarizes also leads to information processing devices, such as spatial light modulators based on liquid crystals~\cite{deGennes1995}. Indeed analogies between spintronics and optics led to the emerging field of spinoptronics~\cite{Shelykh2004}, aiming to exploit a hybridization of the different fields.

This hybridization is well illustrated by semiconductor microcavities in the strong coupling regime, which have become a promising basis for all-optical devices and circuits~\cite{Liew2011,EspinosaOrtega2013,AmoNatPhot,Cerna,amobullet}. The strong coupling regime gives rise to exciton-polaritons (for simplicity, polaritons), whose light mass and integer spin facilitates condensation at elevated temperatures~\cite{Christopoulos2007,KenaCohen2010}. The excitonic fraction leads to strong carrier-carrier interactions~\cite{Savvidis2000} and sensitivity to gating electromagnetic fields \cite{TsotsisPRA}. At the same time, the photonic fraction allows the spin state of the polariton to be directly imprinted onto the polarization of the emitted light. An advantage over conventional spintronic devices is that, being neutral particles, polaritons do not experience strong dephasing due to Coulomb scattering and the coherent propagation of spin currents can be achieved over hundreds of microns~\cite{Kammann}.

Optically controlled carrier-carrier interactions in strongly coupled semiconductor microcavities have been shown to be a highly effective tool in the engineering of polaritons' energy landscapes~\cite{Amo2010}, for both the study of polariton condensate phenomena~\cite{Wertz, Tosi, ChristmannPRB, ChristofoliniPRL} and the realization of nascent polariton devices~\cite{Gao, AntonAPL, ballarini, Nguyen, AntonPRB,sturm}. The interaction strength is spin dependent~\cite{vina96,Glazov}; excitons with parallel spins experience a repulsive force, while the interaction between excitons with anti-parallel spin is weaker and can be attractive in nature~\cite{Vladimirova}.
\begin{figure}
\setlength{\abovecaptionskip}{-5pt}
\setlength{\belowcaptionskip}{-2pt}
\begin{center}
\includegraphics[width=1\linewidth,angle=0]{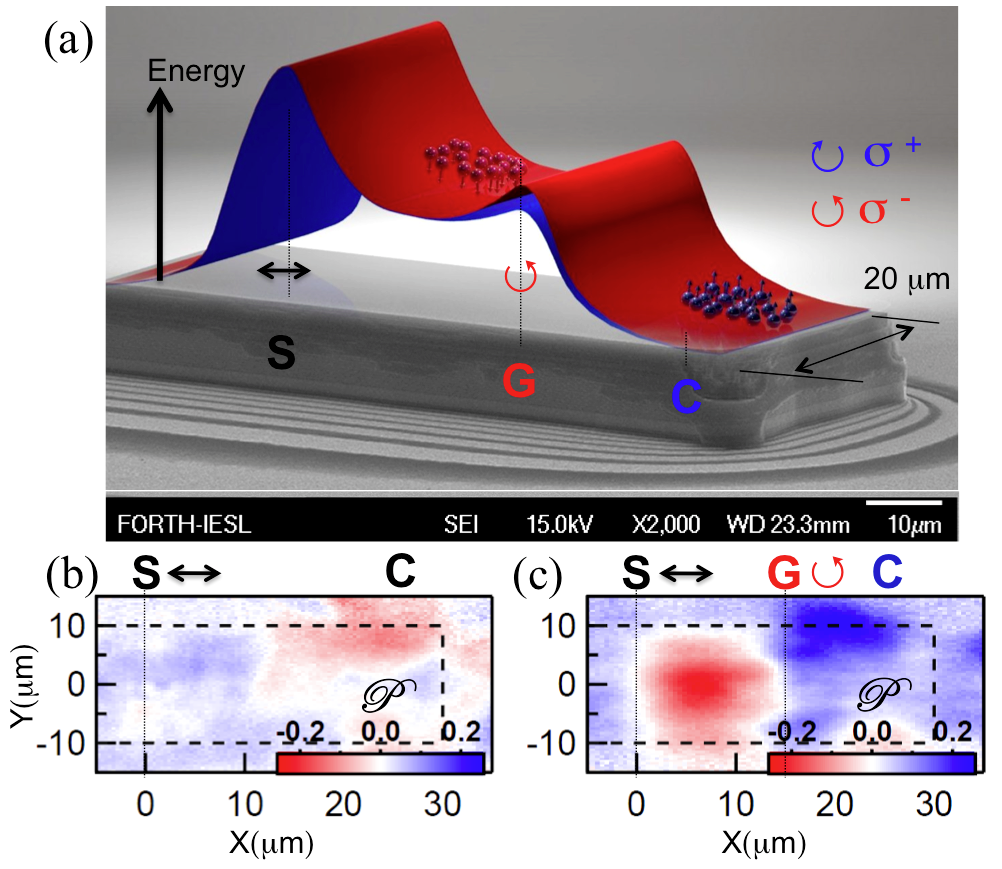}
\end{center}
\caption{(Color online) (a) A schematic of the spin filter with the red and blue slopes depicting the energy landscape for $\sigma^+$ and $\sigma^-$ polaritons for a $\sigma^-$ ($\circlearrowleft$) gate (G). The source beam (S) has linear polarization. The degree of circular emission at C in real space (b) without and (c) with G present where the black dashed line shows the region of interest of the ridge.}
\label{fig:schematic}
\end{figure}
\begin{figure*}
\setlength{\abovecaptionskip}{-5pt}
\setlength{\belowcaptionskip}{-2pt}
\begin{center}
\includegraphics[width=1\linewidth,angle=0]{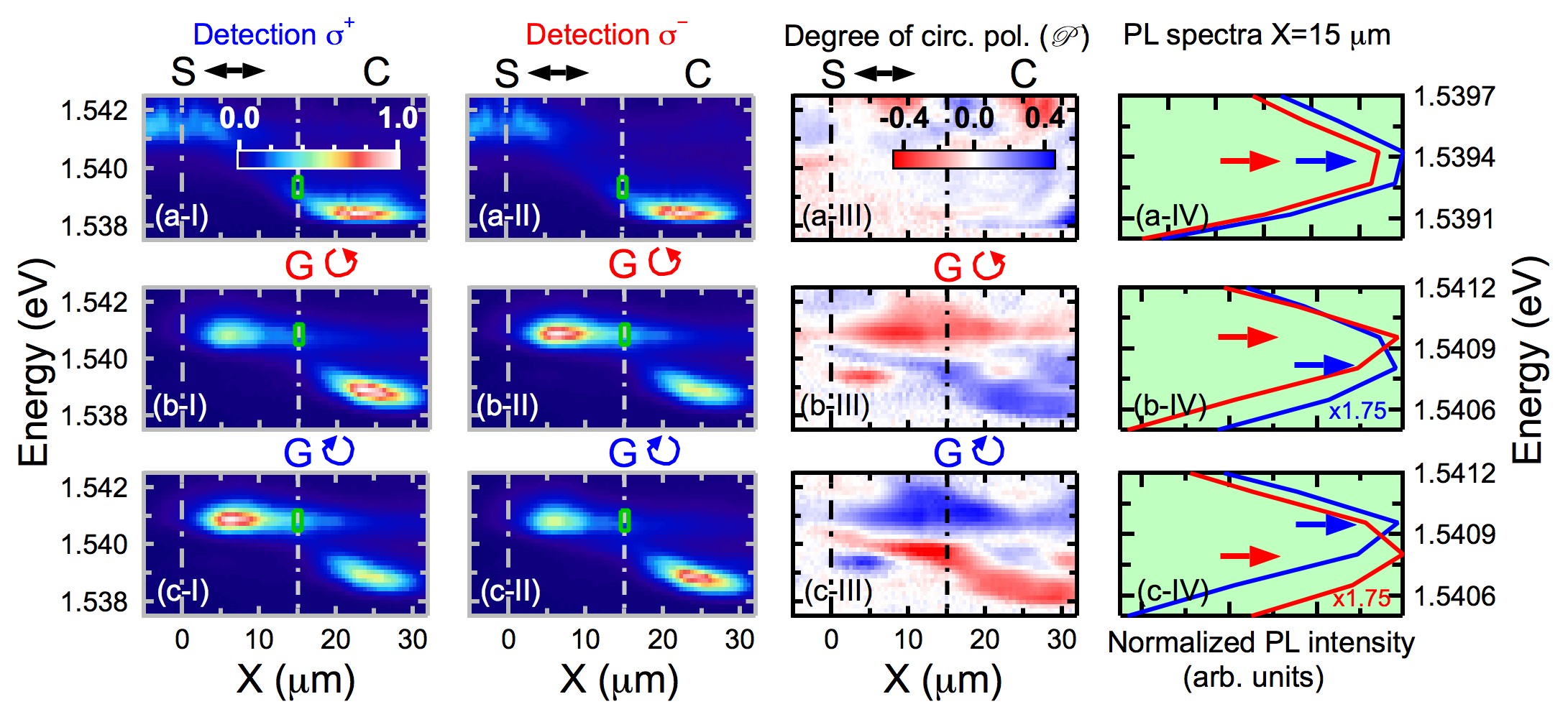}
\end{center}
\caption{(Color online) PL intensity emitted along the center of the ridge as function of energy and real-space (X), under non-resonant, CW excitation for different S and G configurations: (a) Linearly polarized ($\leftrightarrow$) S only, (b) linearly polarized S and $\sigma^-$ polarized ($\circlearrowleft$) G, (c) linearly polarized S and $\sigma^+$ polarized ($\circlearrowright$) G. Columns (I) and (II) show the polariton PL under $\sigma^+$ and $\sigma^-$ detection, respectively; for each row (a-c), the intensity has been normalized to the maximum value obtained at a given detection. Column (III) depicts the degree of circular polarization ($\mathscr{P}$). Column (IV) zooms the normalized PL intensity emitted at $X=$15 $\mu$m (G spot position) under $\sigma^+$ (blue) and $\sigma^-$ (red) detection as function of energy in a small range, indicated by the green boxes in columns (I-II); to evidence the energy splitting (see blue/red arrow pointing at the spectrum peak $E_{\sigma^+}$/$E_{\sigma^-}$) the spectrum for $\sigma^+$/$\sigma^-$ detection in panel (b-IV)/(c-IV) is multiplied by $1.75$. The S/G power is 5.7 $P_{th}$ / 0.7$P_{th}$.}
\label{fig:panels}
\end{figure*}

In this work we show that these anisotropic interactions allow the construction of a photonic analogue of current spin polarization as in a ferromagnet. Namely, we demonstrate spin polarization control of a polariton condensate signal using an optical gate in a high-finesse microcavity ridge. A schematic of the device is shown in Fig. \ref{fig:schematic}(a) where the optically-imprinted potential landscapes for $\sigma^-$ ($\circlearrowleft$, red) and $\sigma^+$ ($\circlearrowright$, blue) polaritons are superimposed onto an SEM image of the microcavity ridge. To produce these landscapes a linearly polarized ($\leftrightarrow$) source beam (S) with a power greater than the polariton condensation threshold ($P_{th}$) injects carriers into the ridge increasing the potential energy at S. This carrier-induced blueshift accelerates polaritons along the ridge, which, if unimpeded, will propagate to the collector (C) [see Fig. \ref{fig:schematic}(a)]. The linear polarization of S ensures equal energy blueshifts for both circularly-polarized polariton states. This is evidenced by a photoluminiscence (PL) whose degree of circular polarization, $\mathscr{P}$, is close to zero when only S is present [Fig. \ref{fig:schematic}(b)]. $\mathscr{P}$ is defined as $(I_{\sigma^+}-I_{\sigma^-})/(I_{\sigma^+}+I_{\sigma^-})$, where $I_{\sigma^+}$ ($I_{\sigma^-}$) is the PL for $\sigma^+$ ($\sigma^-$) detection. A second $\sigma^-$-polarized gate beam (G) produces unbalanced spin-polarized populations of the photo-generated carriers and therefore spin-dependent blueshifts at G. The different potential barriers seen by propagating $\sigma^+$ and $\sigma^-$ polaritons, corresponding to an effective Zeeman splitting, lead to a more efficient blocking of $\sigma^-$ polaritons and thus a net polarization of condensed polaritons between S and G is obtained. This polarization control is evidenced by nonzero $\mathscr{P}$ values when G is present [Fig. \ref{fig:schematic}(c)] \cite{Footnote1}.

The experiments are conducted on a ridge 300 $\times$ 20 $\mu$m$^2$ formed via reactive ion etching of a high-finesse planar microcavity (Q$>$16000) \cite{Gao,Tsotsis}. The sample is placed into a cryostat cooled to 30 K and excited non-resonantly using a microscope objective (NA=0.55) producing excitation spots of $\sim$2 $\mu$m-diameter. The PL is collected with the same objective and analyzed using an imaging spectrometer to resolve the emission simultaneously in real space and energy. A combination of waveplates and polarizers is used to measure the degree of circular polarization of the PL. The laser is mechanically chopped (duty cycle 5\%) to minimize heating of the ridge. S and G are kept close to the end of the ridge in order to focus on the spin filtering operation rather than spin transport. 

The energy-resolved PL is collected along the long axis of the ridge ($X$) and from its center ($Y=0$,$\Delta Y=5\mu$m) for different configurations of G, under a linear-polarized S, see Fig. \ref{fig:panels}. Column I (II) shows normalized PL maps under $\sigma^+$ ($\sigma^-$) detection. Column III compiles $\mathscr{P}$. In column IV, blue and red lines ($\sigma^+$ and $\sigma^-$ detection, respectively) depict the PL spectra at G, energy-zoomed as indicated by the small, green boxes in columns I and II.

Figures \ref{fig:panels}(a-I-IV) compile the results under excitation with only the linearly polarized S (5.7$P_{th}$). In both Figs. \ref{fig:panels}(a-I) and \ref{fig:panels}(a-II) emission is observed at S ($X=0$) and C ($X\approx25$ $\mu$m). The difference between the energy of the emission at S  ($\sim$1.541 eV) and C ($\sim$1.538 eV) originates from the carrier induced blueshift at S. Similarity between Figs. \ref{fig:panels}(a-I) and \ref{fig:panels}(a-II) is illustrated in Fig. \ref{fig:panels}(a-III) which shows no net circular polarization of the emission from the ridge. As shown in Fig. \ref{fig:panels}(a-IV), in the absence of G, both spectra peak at the same energy ($\Delta E =E_{\sigma^+}-E_{\sigma^-}=0$) at $X=15$ $\mu$m.

The corresponding results when an additional $\sigma^-$-polarized, below threshold (0.7$P_{th}$) G beam is introduced between S and C (at $X=15$ $\mu$m) is illustrated in Figs. \ref{fig:panels}(b-I-IV). The PL between S and G ($X\approx5$ $\mu$m) arises from polaritons stopped by the potential barrier at G. There is a striking difference between the maps shown in Figs. \ref{fig:panels}(b-I) and \ref{fig:panels}(b-II): in the former case, the PL at C is stronger than that of the trapped S-G state, while in the latter one, the opposite situation is observed. This preferential blocking of $\sigma^-$ polaritons induces a positive degree of circular polarization in the PL from C and a negative one from the trapped S-G condensate, as seen in Fig. \ref{fig:panels}(b-III). In Fig. \ref{fig:panels}(b-IV), the zoomed spectra at G show polariton distributions with a peak at higher energy for $\sigma^-$ polaritons ($\Delta E<0$). 

The reversibility of this effect is observed in Figs. \ref{fig:panels}(c-I-IV), where the polarization of G is set to $\sigma^+$. The change of sign in the energy splitting, $\Delta E>0$, demonstrates the spin dependence of the blueshifts on the G polarization.

%Up to here changes accepted

%
\begin{figure}
\setlength{\abovecaptionskip}{-5pt}
\setlength{\belowcaptionskip}{-2pt}
\begin{center}
\includegraphics[width=1\linewidth,angle=0]{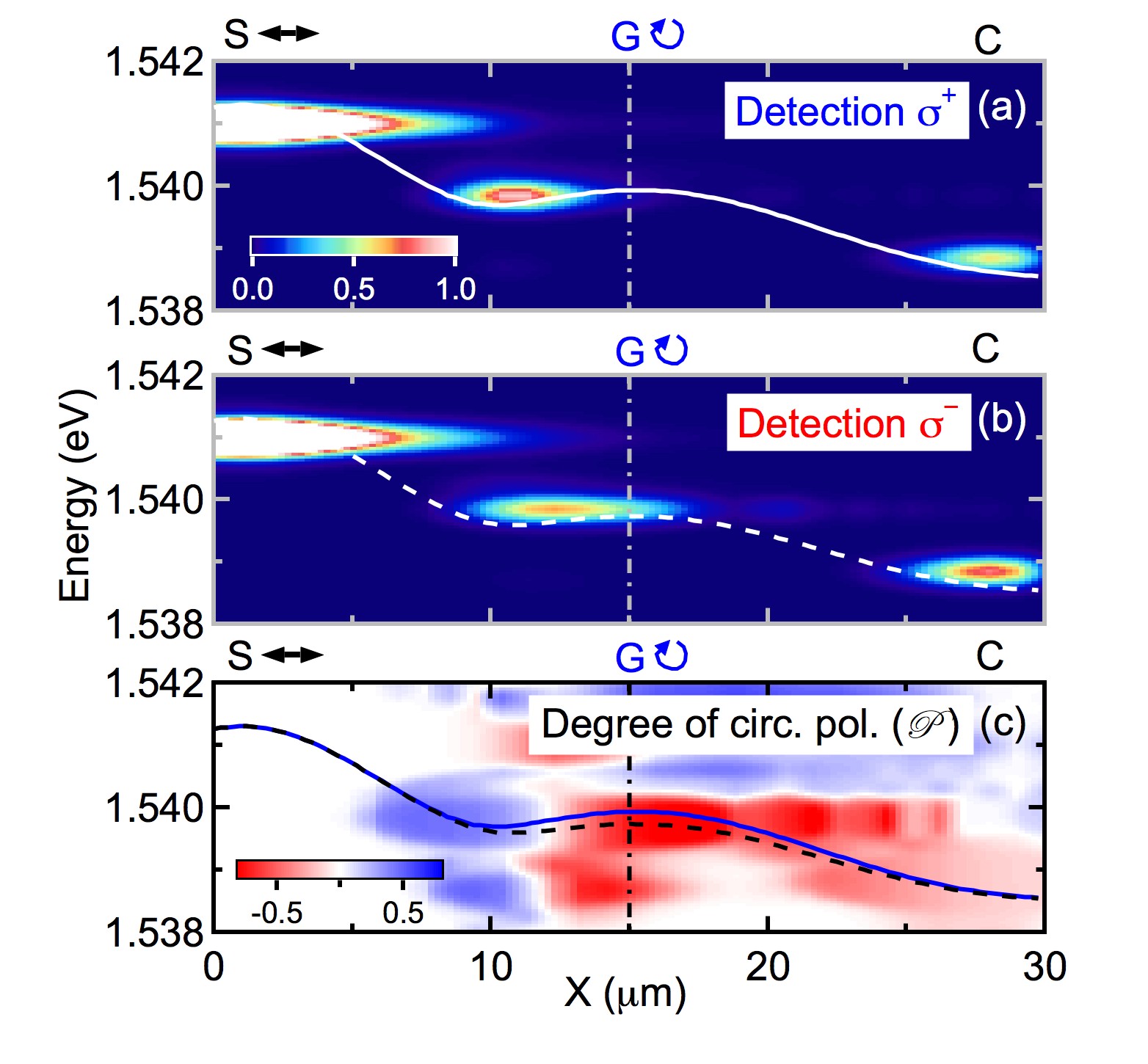}
\end{center}
\caption{(Color online) (a) [(b)] Simulation of the PL intensity, under $\sigma^+$ ($\sigma^-$) detection, $|\psi(\mathbf{x})_+|^2$ ($|\psi(\mathbf{x})_-|^2$) emitted at the center of the ridge as function of energy and real-space (X) for linearly polarized S ($\leftrightarrow$) and $\sigma^+$ polarized ($\circlearrowright$) G, obtained theoretically from Eq.~\ref{eq:GP}. In panels (a,b) the intensity has been normalized to the maximum PL value of the condensate stopped by G under $\sigma^+$ detection. (c) Degree of circular polarization ($\mathscr{P}$) obtained from previous panels. In panels (a-c), the effective potentials $V_\pm(\mathbf{x})$ experienced by polaritons of different spin polarizations are shown in solid and dashed lines, respectively. The simulated PL intensity and $\mathscr{P}$ maps are coded in false, linear color scales.}

\label{fig:theory}
\end{figure}

The spatial structure of polariton condensates is typically described using a mean-field description, generalized to include incoherent pumping and decay~\cite{WoutersPRL}. The incoherent pumping excites a hot exciton reservoir, which can be assumed to have a steady density profile in a continuous wave experiment. Excitons then undergo stimulated relaxation into the polariton condensate, which can be described using the Landau-Ginzburg approach~\cite{Keeling2008}, here generalized to account for the spin degree of freedom and energy relaxation~\cite{WoutersNJP}. The evolution of the 2D spinor polariton wavefunction $\psi_\sigma(\mathbf{x},t)$ is:
\begin{align}
i\hbar&\frac{d\psi_\sigma(\mathbf{x},t)}{dt}=\left[\hat{E}_{LP}+\left(\alpha_1-i\Gamma_\mathrm{NL}\right)|\psi_\sigma(\mathbf{x},t)|^2\right.\notag\\
&\hspace{5mm}\left.+\alpha_2|\psi_{-\sigma}(\mathbf{x},t)|^2+V_\sigma(\mathbf{x})+iW_\sigma(\mathbf{x})-\frac{i\Gamma}{2}\right]\psi_\sigma(\mathbf{x},t)\notag\\
&\hspace{5mm}+i\hbar\mathfrak{R}\left[\psi(\mathbf{x},t)\right].\label{eq:GP}
\end{align}
where $\sigma=\pm$ denotes the two circular polarizations of polaritons.  $\hat{E}_{LP}$ represents the kinetic energy dispersion of polaritons, which at small wavevectors can be approximated as $\hat{E}_{LP}=-\hbar^2\hat{\nabla}^2/\left(2m\right)$, where $m$ is the polariton effective mass. $\alpha_1$ and $\alpha_2$ represent the strengths of interactions between polaritons with parallel and antiparallel spins, respectively. Polaritons enter the condensate at a rate determined by $W_\sigma(x,t)$, which is both polarization and spatially dependent. While the non-resonant laser used in the experiment is polarized, it in general excites both spin polarizations due to the partial spin relaxation of hot excitons during their relaxation to form polaritons. The condensation rate is then given by:
\begin{equation}
W_\sigma(\mathbf{x})=P_\sigma(\mathbf{x})+rP_{-\sigma}(\mathbf{x})
\end{equation}
where $P_\sigma(\mathbf{x})$ is the spatial profile of the pump intensity and $r$ is a phenomenological constant. It is implicit that this form also includes any spin anisotropy in the condensation rates.

In addition to driving the polariton condensate, the hot exciton reservoir also provides a spin dependent effective potential for polaritons:
\begin{equation}
V_\sigma(x)=V_0(\mathbf{x})+G\left[P_\sigma(\mathbf{x})+rP_{-\sigma}(\mathbf{x})\right]^p=V_0(\mathbf{x})+GW^p_\sigma(\mathbf{x})
\end{equation}
where $G$ is a constant representing the strength of forward scattering processes between excitons in the reservoir and in the condensate. We have assumed that any spin anisotropy of these processes is the same as the spin anisotropy in the scattering of excitons from the reservoir into the polariton condensate. In addition, our experimental measurements showed that the effective potential increases sub-linearly with the pump power. For this reason, we introduce the constant $p$, which can be obtained empirically. $V_0$ represents a spin-independent component to the effective potential, which represents the walls of the ridge and non-uniform potential along the ridge. In particular, previous studies showed that there is a slight drop of the polariton potential near the end of the ridge~\cite{AntonPRB,AntonPRBspeed}.

The polaritons decay with a decay rate $\Gamma$ and also experience a nonlinear loss $\Gamma_\mathrm{NL}$ corresponding to scattering out of the condensate~\cite{Keeling2008}.

The final term in Eq.~\ref{eq:GP} accounts for energy relaxation processes of condensed polaritons:
\begin{equation}
\mathfrak{R}[\psi(x,t)]=-\left(\nu+\nu^\prime|\psi(x,t)|^2\right)\left(\hat{E}_\mathrm{LP}-\mu(x,t)\right)\psi(x,t),\label{eq:relax1}
\end{equation}
where $\nu$ and $\nu^\prime$ determine the strength of energy relaxation~\cite{WoutersNJP,WertzPRL,Solnyshkov2013,Sieberer2013} and $\mu(x,t)$ is a local effective chemical potential that conserves the polariton population~\cite{WoutersNJP,AntonPRB}. The terms cause the relaxation of any kinetic energy of polaritons and allow the population of lower-energy states trapped between the pump-induced potentials.

Solving Eq.~\ref{eq:GP} numerically	\cite{Parameters} gives the results shown in Fig.~\ref{fig:theory}, which can be compared to the experimental results in Fig.~\ref{fig:panels}. The solid and dashed curves show the polariton effective potentials $V_\sigma(\mathbf{x})$. Although G only induces a slight difference in the potentials for the two spin components, it is enough to preferentially block the passage of polaritons co-polarized to G, while C polaritons have an opposite polarization to that of G.

\begin{figure}
\setlength{\abovecaptionskip}{-5pt}
\setlength{\belowcaptionskip}{-2pt}
\begin{center}
\includegraphics[width=1\linewidth,angle=0]{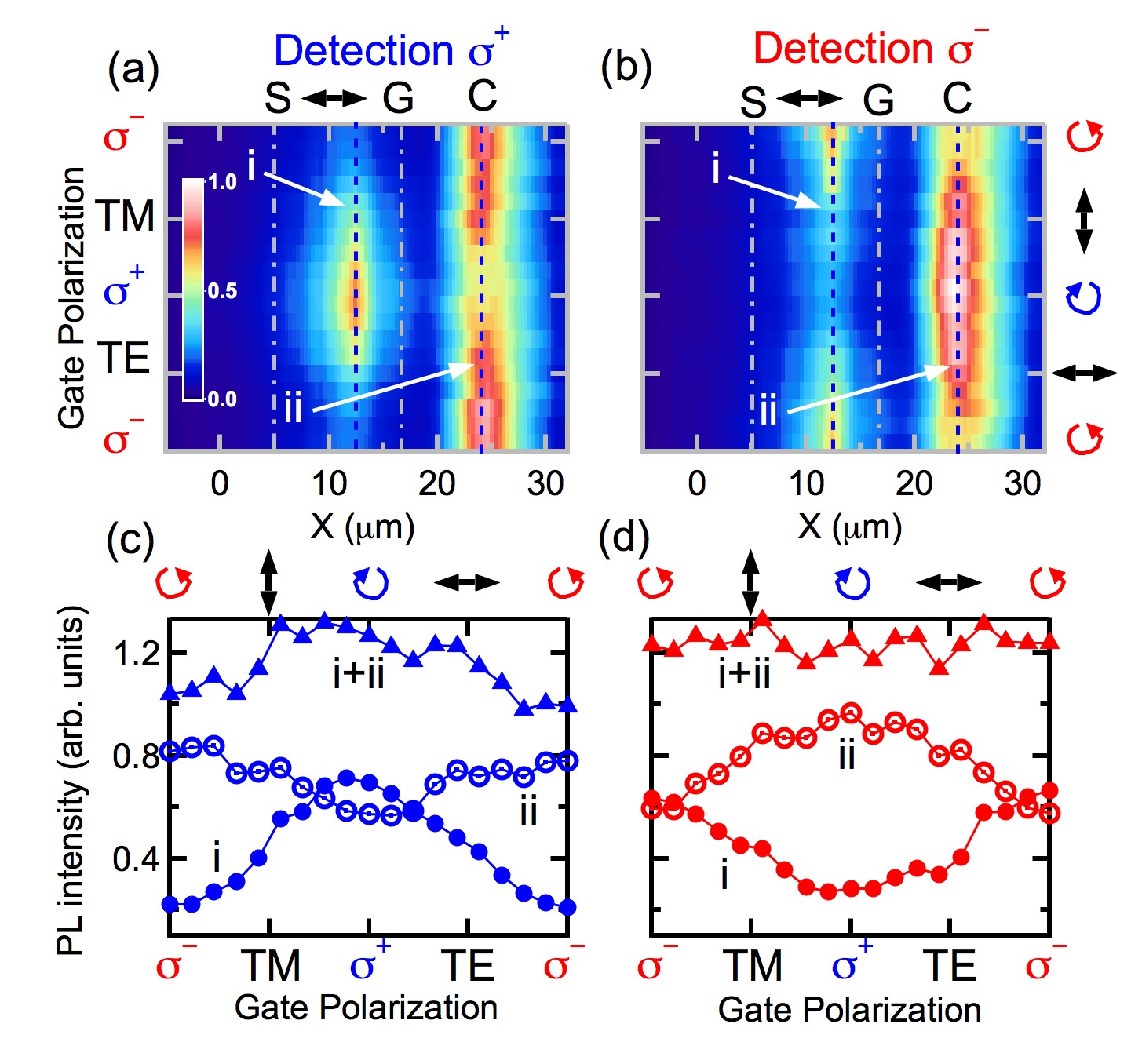}
\end{center}
\caption{(Color online) Spin filtering dependence on the polarization of G. (a)/(b) PL intensity map under $\sigma^+$/$\sigma^-$ detection emitted at the center of the ridge as function of G polarization and real-space (X), S is linearly polarized ($\leftrightarrow$) and the polarization of G is continuously varied from $\sigma^-$ to $\sigma^+$ (vertical axis). The intensity maps have been normalized to the maximum PL value of the condensate stopped at C under $\sigma^-$ detection. Dot dash, vertical lines indicate the positions of S and G, at 5 and 16 $\mu$m, respectively. Dashed, vertical lines, dubbed as $i$ and $ii$ (at $X\approx$12 and 25 $\mu$m, respectively), mark the cross-section plotted in panels (c,d). (c)/(d) PL intensity at the cross-sections $i$ and $ii$ as function of G polarization in full and open circles, respectively. The horizontal trace $i+ii$ (full triangles) is the result of adding $i$ and $ii$.}
\label{fig:Tuning}
\end{figure}

We now investigate the control of the spin of the condensate at C by varying the polarization of G. Figure \ref{fig:Tuning} shows the PL at the center of the ridge ($Y=0$) and along its $X$ axis, for $\sigma^+$ (a) and  $\sigma^-$ (b) detection as the polarization of G is changed (vertical axis). It should be mentioned that now the S beam is located at 5 $\mu$m, instead at $X=0$, as it was the case in Figs. \ref{fig:schematic}-\ref{fig:theory}. In Fig. \ref{fig:Tuning}(a)/(b), $\sigma^+$/$\sigma^-$ detection, the PL at C is maximum/minimum for a $\sigma^-$-polarized G, confirming the spin-selective filtering of the polarized G. As G becomes linearly polarized (TM or TE), similar emission is observed at C in both Fig. \ref{fig:Tuning}(a) and \ref{fig:Tuning}(b). As the polarization of G reaches $\sigma^+$ the emission from C is minimized/maximized for $\sigma^+$/$\sigma^-$ detection [Fig. \ref{fig:Tuning}(a)/(b)]. Figs. \ref{fig:Tuning}(c) and \ref{fig:Tuning}(d) plot the PL intensities at the S-G trapped condensate ($i$, $X=12$ $\mu$m), at C ($ii$, $X=25$ $\mu$m) and their sum ($i+ii$), for $\sigma^+$ and $\sigma^-$ detection, respectively. Under $\sigma^+$ detection, it is clearly seen, Fig. \ref{fig:Tuning}(c), that $\sigma^+$-polaritons are efficiently blocked by G when its polarization is $\sigma^+$, leading to the peak observed in trace $i$ (full circles); concomitantly a dip is observed for these conditions in trace $ii$ (open circles). The reverse situation holds for a $\sigma^-$-polarized G when a minimum/maximum occurs in curve $i$/$ii$. The constant value of the addition $i+ii$ (up triangles) reflects the fact that there are not significant losses of polaritons traveling from S to C. In Fig. \ref{fig:Tuning}(b) [Fig. \ref{fig:Tuning}(d)], under $\sigma^-$ detection, the results are equivalent to those described in Fig. \ref{fig:Tuning}(a) [Fig. \ref{fig:Tuning}(c)] by interchanging $\sigma^-\leftrightarrow\sigma^+$ and TE$\leftrightarrow$TM in the vertical [horizontal] axis. The maxima for the S-G condensates [$i$ traces in Figs. \ref{fig:Tuning}(c,d)] are obtained for the same polarization of G as that of the detection; a similar situation is found for the minima of the C condensates ($ii$ traces). 

The modulation in the S-G and C populations, induced by the polarization of G, is limited by the difference between the spin-dependent blueshifts. Using a G beam resonant with the exciton reservoir should allow a larger modulation. These modulations observed in both polarization detections are analogous to the oscillations of the current in spin transistors \cite{DataDas, Shelykh} however the spin current is controlled by a gate with constant intensity.

 In conclusion we have demonstrated the ability to control the PL polarization of a polariton condensate with a low-power optical gate. The ability to optically control the polarization of polariton fluxes using spin dependent potential energy landscapes promises new functionality for polariton and excitonic devices. For example a polariton spin flux optical router should be possible in a cross shaped ridge utilizing a similar scheme to the successful optical routing of exciton flux in coupled quantum wells \cite{Andreakou}.

We acknowledge EU ITN grant INDEX 289968, Greek GSRT ARISTEIA program Apollo and Spanish MINECO MAT2011-22997. C.A. acknowledges financial support from a Spanish FPU scholarship. PS  gratefully acknowledges financial support from the Leverhulme Trust.


\begin{thebibliography}{}
\bibitem{Footnote1} {The lower values of $\mathscr{P}$ in figs. \ref{fig:schematic}(b,c) compared to column 3 of fig. 2 is because the polariton emission is integrated in energy (grating at 0 nm) leading to lower values of polarization (between -0.2 and 0.2).}
\end{thebibliography}
\end{document}